 \documentclass[preprint2]{aastex631}

\usepackage{subfigure}
\usepackage{graphicx}
\usepackage{amsmath}
\usepackage{bm}

\received{2023 May 23}
\accepted{2023 July 6}

\submitjournal{ApJ}

\shorttitle{Superflares on EI~Cnc}
\shortauthors{Li et al.}

\begin{document}

\title{White-light superflare and long-term activity of the nearby M7 type binary  EI~Cnc observed with GWAC system}

\correspondingauthor{Huali Li, Jing Wang}
\email{lhl@nao.cas.cn, wj@nao.cas.cn}

\author{Hua-Li Li}
\affiliation{CAS Key Laboratory of Space Astronomy and Technology, National Astronomical Observatories, Chinese Academy of Sciences, Beijing 100101, China.}

\author{Jing Wang}
\affiliation{Guangxi Key Laboratory for Relativistic Astrophysics, School of Physical Science and Technology, Guangxi University, Nanning 530004, China}
\affiliation{CAS Key Laboratory of Space Astronomy and Technology, National Astronomical Observatories, Chinese Academy of Sciences, Beijing 100101, China.}

\author{Li-Ping Xin}
\affiliation{CAS Key Laboratory of Space Astronomy and Technology, National Astronomical Observatories, Chinese Academy of Sciences, Beijing 100101, China.}

\author{Jian-Ying Bai}
\affiliation{CAS Key Laboratory of Space Astronomy and Technology, National Astronomical Observatories, Chinese Academy of Sciences, Beijing 100101, China.}

\author{Xu-Hui Han}
\affiliation{CAS Key Laboratory of Space Astronomy and Technology, National Astronomical Observatories, Chinese Academy of Sciences, Beijing 100101, China.}

\author{Hong-Bo Cai}
\affiliation{CAS Key Laboratory of Space Astronomy and Technology, National Astronomical Observatories, Chinese Academy of Sciences, Beijing 100101, China.}

\author{Lei Huang}
\affiliation{CAS Key Laboratory of Space Astronomy and Technology, National Astronomical Observatories, Chinese Academy of Sciences, Beijing 100101, China.}

\author{Xiao-Meng Lu}
\affiliation{CAS Key Laboratory of Space Astronomy and Technology, National Astronomical Observatories, Chinese Academy of Sciences, Beijing 100101, China.}

\author{Yu-Lei Qiu}
\affiliation{CAS Key Laboratory of Space Astronomy and Technology, National Astronomical Observatories, Chinese Academy of Sciences, Beijing 100101, China.}

\author{Chao Wu}
\affiliation{CAS Key Laboratory of Space Astronomy and Technology, National Astronomical Observatories, Chinese Academy of Sciences, Beijing 100101, China.}

\author{Guang-Wei Li}
\affiliation{CAS Key Laboratory of Space Astronomy and Technology, National Astronomical Observatories, Chinese Academy of Sciences, Beijing 100101, China.}

 \author{Jing-Song Deng}
 \affiliation{CAS Key Laboratory of Space Astronomy and Technology, National Astronomical Observatories, Chinese Academy of Sciences, Beijing 100101, China.}
 \affiliation{School of Astronomy and Space Science, University of Chinese Academy of Sciences, Beijing, China}

\author{Da-Wei Xu}
\affiliation{CAS Key Laboratory of Space Astronomy and Technology, National Astronomical Observatories, Chinese Academy of Sciences, Beijing 100101, China.}
\affiliation{School of Astronomy and Space Science, University of Chinese Academy of Sciences, Beijing, China}

  \author{Yuan-Gui Yang}
  \affiliation{School of Physics and Electronic Information, Huaibei Normal University, Huaibei 235000, China. }

\author{Xiang-Gao Wang}
\affiliation{Guangxi Key Laboratory for Relativistic Astrophysics, School of Physical Science and Technology, Guangxi University, Nanning 530004, China}

\author{En-Wei Liang}
\affiliation{Guangxi Key Laboratory for Relativistic Astrophysics, School of Physical Science and Technology, Guangxi University, Nanning 530004, China}

\author{Jian-Yan Wei}
\affiliation{CAS Key Laboratory of Space Astronomy and Technology, National Astronomical Observatories, Chinese Academy of Sciences, Beijing 100101, China.}



\begin{abstract}
Stellar white-light flares are believed to play an essential role on the physical and chemical properties of the atmosphere of the surrounding exoplanets.   
Here we report an optical monitoring campaign on the nearby flaring system 
EI~Cnc carried out by the Ground-based Wide Angle Cameras (GWAC) and its dedicated follow-up telescope. 
A superflare, coming from the brighter component EI~CncA, was detected and observed,
in which four components are required to properly model the complex decay light curve.
The lower limit of flare energy in the $R-$band is estimated to be $3.3\times10^{32}$ ergs.
27 flares are additionally detected from the GWAC archive data with a total duration of 290 hours. The inferred cumulative flare frequency distribution follows a quite shallow 
power-law function with a slope of $\beta=-0.50\pm 0.03$ over the energy range between $10^{30}$ and 
$10^{33}$ erg, which reinforces the trend that stars cooler than M4 show enhanced
superflare activity. The flares identified in EI~Cnc enable us to extend the 
$\tau-E$ relationship previously 
established in the white-light superflares of solar-type stars down to an energy as low as
$\sim10^{30}$erg (i.e., by three orders): $\tau\propto E^{0.42\pm0.02}$, 
which suggests a common flare mechanism for stars with a type from M to solar-like,
and implies an invariant of $B^{1/3}\upsilon_{\rm A}$
in the white-light flares.
\end{abstract}

\keywords{flare --- stars: individual (EI~Cnc)---techniques: photometric--- techniques:}

\section{Introduction}           
\label{sect:intro}

M dwarfs occupy a majority ($\sim 70-75\%$) 
of the local stellar population \citep{Chabrier1997,Henry2006,Bochanski2010}.
A large fraction of M dwarfs are magnetically active with flares more energetic than those of the Sun \citep{Hawley2014}.
Compare to solar-like and early-M stars,
the understanding of the dynamo pattern is a challenge for the ultra-cool dwarfs (UCDs) \citep{Chabrier1997,Mohanty2002}, 
because the solar-type shell dynamo is hard to be supported in UCDs due to the lack of the tachocline - a narrow boundary layer separating the convective and radiative zones \citep{Mohanty2002,Charbonneau2014,Kochukhov2021}.

Flares from UCDs have been ubiquitously observed in multi-wavelength, including near-infrared
\citep{Kanodia2022}, optical bands
\citep[e.g.,][]{Fuhrmeister2004a,Stelzer2006,Schmidt2014,Schmidt2016,Gizis2017,Paudel2018,Xin2021a,Xin2023a},
ultraviolet, and X-ray \citep[e.g.,][]{Fleming2000,Stelzer2006, Robrade2010, Deluca2020}.
\citet{Gizis2000} estimated a flaring frequency of 7$\%$ or higher during the whole life of the late-type stars with the studies of  a sample of 53 nearby M7-M9.5 dwarfs, suggesting that flaring is common among ultracool dwarfs. In addition, \citet{Hawley2014} estimated that the flaring time of active and inactive stars varies between $\sim30\%$ and $\sim0.01\%$ at a level detectable with $Kepler$.
\cite{Paudel2018} reported a total of 283 flares from  
10 UCDs by using the $Kepler$ $K2$ short-cadence data. With a sample of 
1392 flare events,
\citet{Medina2020} found a high value of flaring frequency of 26\% from 125 single 
mid- to late-M dwarfs, in which 
60\% of the sample had flared one or more times. In addition,
\cite{Murray2022} have recently identified 234 flares from 85 flaring UCDs by the SPECULOOS-South survey program, 
suggesting that M5–M7 stars are more likely to flare than both earlier and later M dwarfs.
The flares of UCD is crucial to understand not only how the magnetic energy is converted to gas kinetic energy within the thick convective atmospheres
\citep[e.g.,][]{Barnes2003,Morin2010,McLean2012}, 
but also 
the relation between the activities of stars 
and the habitability of an exoplanet.
Due to their close proximity of habitable zone,
the close-in planetary companions are more likely to be exposed to the high-energy radiation and particles from the hosts \citep[e.g.,][]{Khodachenko2007,Lammer2007,Shields2016}.
On the one hand, 
it is suggested that the ultraviolet radiation and ionized particles
released in flares may initiate chemistry relevant to the origin of life \citep{Ranjan2017, Rimmer2018}. 
On the other hand, it may have effects on the chemical compositions of planetary atmospheres \citep[e.g.,][]{Lammer2007,Segura2010,Tilley2019}. Theoretical studies suggest that
no significant ozone layer destruction occurs, 
if the flares output mainly consists of photons rather than energetic particles \citep[e.g.,][]{Segura2010,Tilley2019, Zeldes2021,Murray2022}.

Here we report an optical monitoring campaign of red dwarf EI~Cnc carried out by the 
ground-based wide angle cameras (GWAC), a ground facility of the SVOM mission (Space-based multi-band astronomical Variable Objects Monitor, \citet{Wei2016}). With a white-light superflare captured by the GWAC in realtime, 
rapid multi-wavelength follow-ups were carried out by a narrow-field optical telescope.
In addition, 27 more flares were identified in the GWAC's archived data. 

The paper is organized as follows. The properties of EI~Cnc and
the used instruments procedure are descripted in Section2. 
Section 3 presents the superflare and the light curve analysis. 
The off-line search of the flares of the object in GWAC archive data is shown in Section 4.  
Section 5 presents the resulted cumulative flare energy distribution.
A discussion is given in Section 6.

\section{The object and observation}

\subsection{EI~Cnc} 

EI~Cnc (G9-38AB, GJ\,1116AB) is a nearby red dwarf system 
whose flaring activity was first reported by \citet{Pettersen1985a}.
This source is located at a distance of about 5.13 pc ($5.136\pm0.003$, $5.126\pm0.005$, \citet{Gaia2018}),
consisting of two nearly identical dwarfs with spectral type of M7+M7 \citep{Newton2014}. 
EI~CncA corresponds to the brighter component with $M_{V}\sim15.46$ mag 
and EI~CncB the fainter one with $M_{V}\sim16.31$ mag \citep{Reid1997}.
The basic properties of both components are listed in Table \ref{Tab1}.
The two components were reported to be separated by $1.39\pm0.01$ arcsec \citep{Law2008}.
Recent measurement given in Gaia DR2 returns a consistent value of $1.37\pm0.01$ arcsec at the epoch of 2015.9 \citep{Gaia2018}. 
By adopting a circular orbit and a total mass of 0.2$M_\odot$, 
the projected separation and orbital period are estimated to be 
7.0 AU and at least 42 years, respectively.

\citet{Pettersen1985b} detected 24 flares in $U$-band from this object within 4.5 hours.
The $U$-band energy released in the flares ranges from $10^{28}$ to $10^{30.5}$ erg.
The object is listed in the nearby flare star catalog complied by \citet{Pettersen1991}, 
and is considered to be slightly less active than the close binary FL~Vir (Wolf 424) \citep{Pettersen2006}.
EI~Cnc is known as a relatively strong X-ray source with $\log L_{\mathrm{X}}/\mathrm{erg\ s^{-1}}=27.5-27.6$
in the ROSAT/PSPC catalog \citep{Schmitt1995,Fleming1995,Schmitt2004}.
A major X-ray flare with a detection of \ion{Fe}{13} coronal line
was reported by \citet{Fuhrmeister2004b}.

\begin{table*}[ht]
   \caption{Properties of EI Cnc extracted from various surveys.}
   \centering
   \begin{tabular}{ccc}
   \hline
    Parameter            & EI~CncA        & EI~CncB     \\
    \hline
    $M_{V}$ (mag)        &   15.46       &       16.31                \\
    \hline
    \multicolumn{3}{c}{Position in Gaia DR2\tablenotemark{1}}    \\
    \hline
    R.A.(2015.5)   &   $134.5593\pm0.09522$   &  $134.55883\pm0.13486$ \\
    Decl.(2015.5)  &   $19.76297\pm0.06141$   &  $19.76258\pm0.08959$  \\
    Parallax (mas)    &    $194.7225\pm0.12507$  &  $195.08365\pm0.17542$ \\
   \hline
   \multicolumn{3}{c}{Spectral classification}        \\
    \hline
    Spectral type       &   M7    &   M7    \\
    $T_{\mathrm{eff}}$ (K)         &       $2905\pm105$ \tablenotemark{2}   &  \dotfill        \\
    \hline
    \multicolumn{3}{c}{Proper motion \tablenotemark{1}}         \\
    \hline                                
    $\mu_{\alpha}$  (mas $\mathrm{yr^{-1}}$) &  $-766.02903\pm0.19043$ & $-938.54588\pm0.25592$           \\ 
    $\mu_{\delta}$  (mas $\mathrm{yr^{-1}}$) &  $-99.26976\pm0.12183$ & $-36.23736\pm0.17225$          \\
    \hline
    \end{tabular}
    \label{Tab1}
    \tablenotetext{1}{\citet{Gaia2021}}
    \tablenotetext{2}{\citet{Wang2022b}}
   \end{table*}

\subsection{The observational facility}

The Ground-based Wide Angle Cameras (GWAC), located at Xinglong observatory of National Astronomical Observatories, Chinese Academy of Sciences, is an array of telescopes with wide-angle cameras dedicated to optical transient survey \citep{Wei2016}.
It monitors the sky with the cadence of 15 seconds (i.e., 10 seconds for exposure and
5 seconds for readout),
aiming to detect short-duration transients such as gamma-ray bursts \citep{Xin2023b}, stellar flares \citep{Wang2021,Wang2022a,Li2023},
optical counterpart of gravitational waves \citep{Turpin2020} and fast radio bursts \citep{Xin2021a}.
Currently, the GWAC is composed of four units.
Each unit is composed of four identical cameras (JFoV) each with an aperture of 
18cm, and one camera (FFoV) with an aperture of 3.5cm that is used to  
capture the bright transients that are saturated in the JFOV. 
The limiting magnitudes of JFoV and FFoV is typical of 16 and 12 magnitude in the $R$-band, respectively. The total field-of-view of the current GWAC is about 2200 square degree.
More details of the GWAC are provided in \citet{Wang2020} and \citet{Han2021}.

\section{The superflare on 2020-02-02}

\subsection{The observations}

\begin{figure}[htbp]
    \centering
    \includegraphics[width=0.3\textwidth]{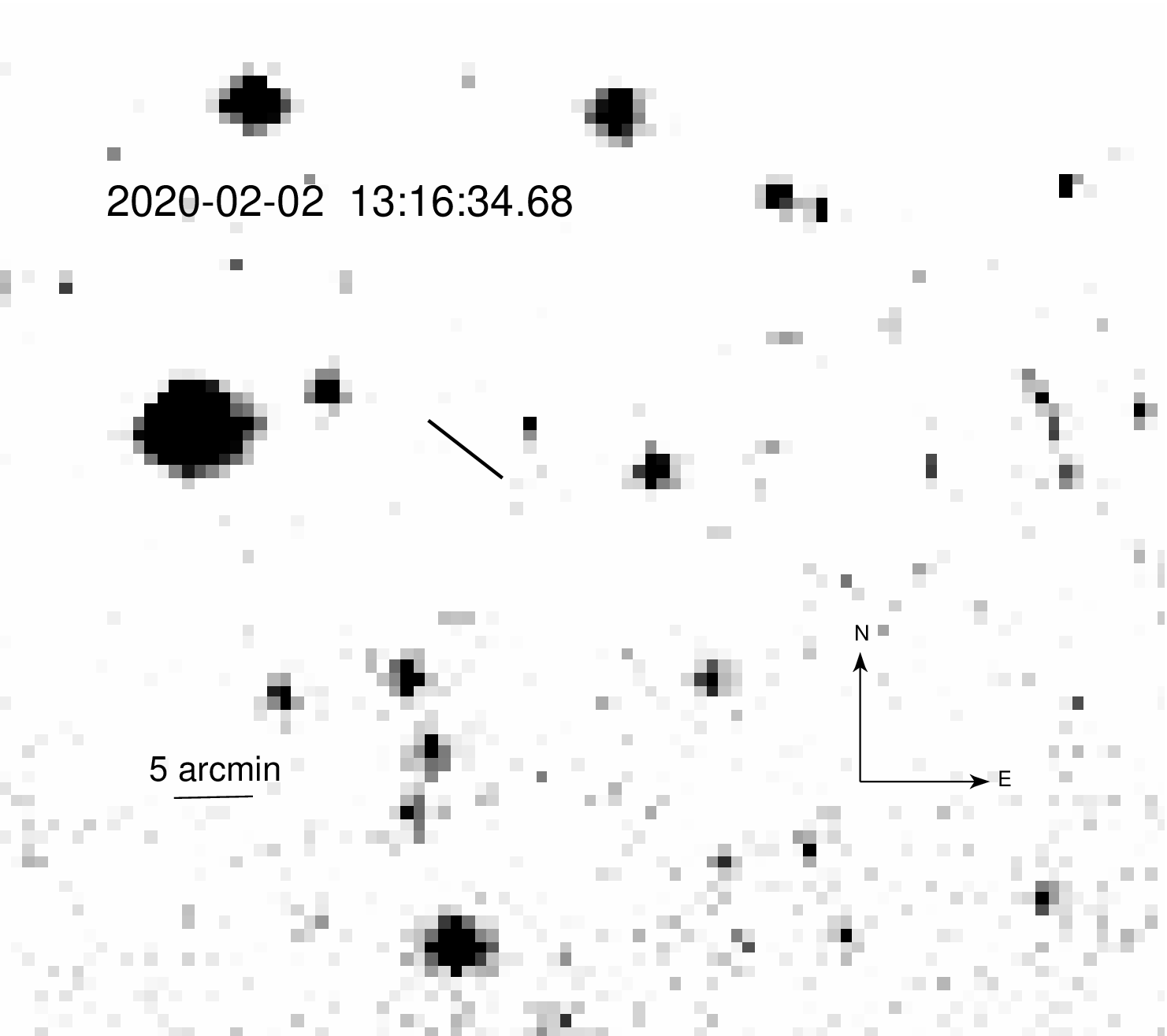} 
   \hspace{.2in} 
   \includegraphics[width=0.3\textwidth]{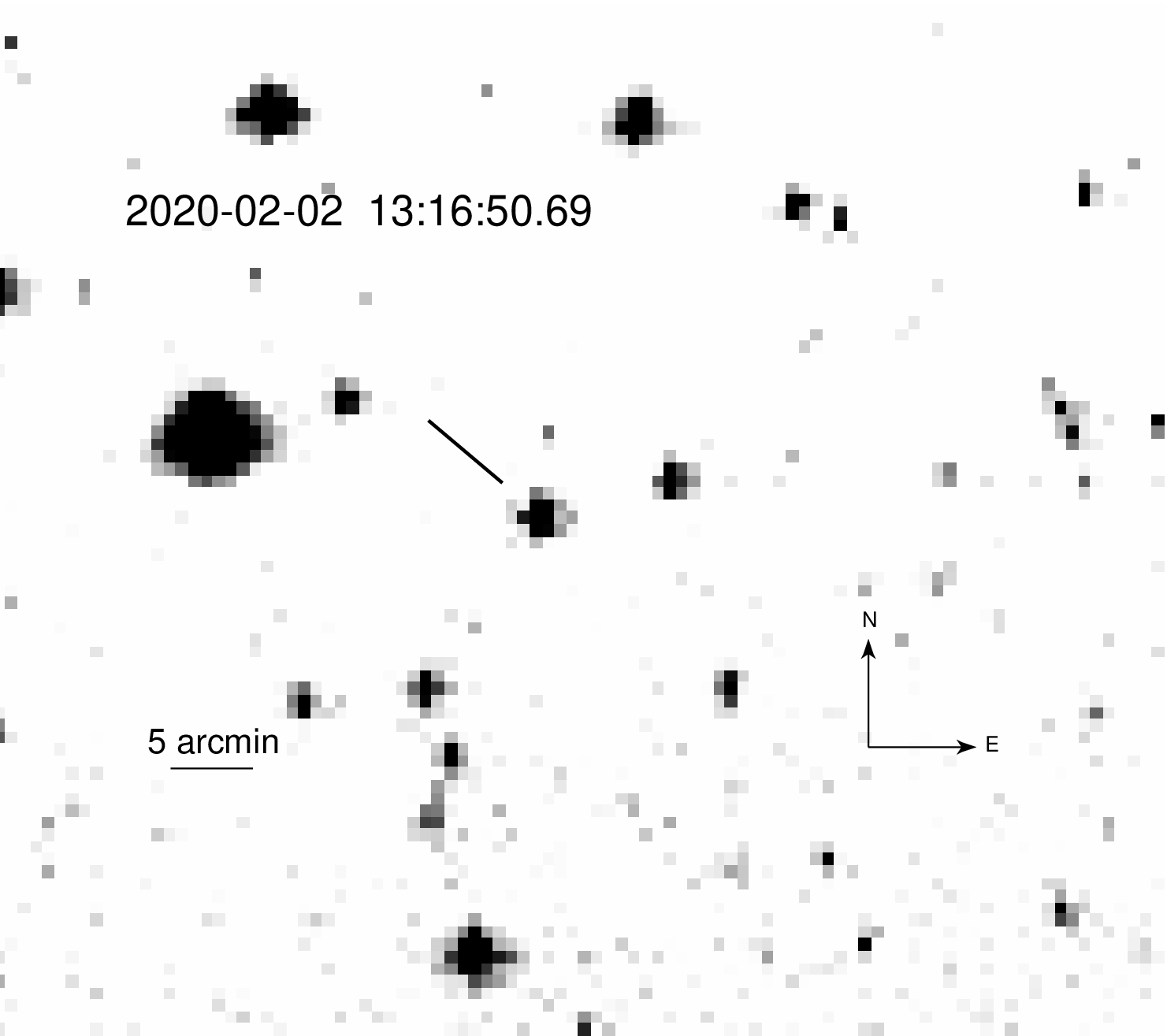}
   \caption{Images of the superflare of EI~Cnc obtained by the GWAC FFoV on 2020-02-02.
   The upper panel is the reference image which was obtained about 15 sec before the triggered of the flare (the lower panel).
   The observation times, direction and spatial scale are labeled in both images.
   }
   \label{fig:fov-GWAC}
   \end{figure}

On 13:16:50 UT Feb. 2, 2020, a white light flare of EI~Cnc was captured in 
real-time by an FFoV camera during its normal survey. 
Figure.\ref{fig:fov-GWAC} displays the reference image (upper panel) and discovery image (lower panel). 
The object was continuously observed by the GWAC from UT12:59:19.6 to 13:31:14.7,  lasting for about half an hour in the field of FFoV on the night.
During the observation, the limiting magnitude of FFoV in the $R-$band is
determined to be about 11.8$\pm$0.2 mag for each single exposure.
The flare decays rapidly to the detection limit of the FFoV 
at 2.5 minutes after the trigger.
Since the observations were continuous with a cadence of 15 seconds, the first brightest measurement is believed to be near the real peak within the temporal resolution, 
which means the brightness was brightened by more than 2.3 magnitudes in 15 seconds during its rising phase\footnote{Such brightening results in a saturation in the JFOV images.}.

A robotic follow-up observation in standard Johnson-Cousins $R$-band was then carried out by the F60A telescope\footnote{The telescope with a diameter of 60cm has a $f$-ratio of 8.0, and is equipped with an Andor 2k$\times$2k CCD. The corresponding pixel scale is 0.52 arcseconds per pixel.}
located beside the 
GWAC cameras via the dedicated real-time automatic transient validation system (RAVS, \cite{Xu2020}).
The follow-up observation started on UTC 13:18:23.5, i.e., 102 seconds after the trigger time of the flare, and lasted for 5.5 hours in total. 

   \begin{figure*}[htbp]
    \centering
       \includegraphics[width=0.4\textwidth]{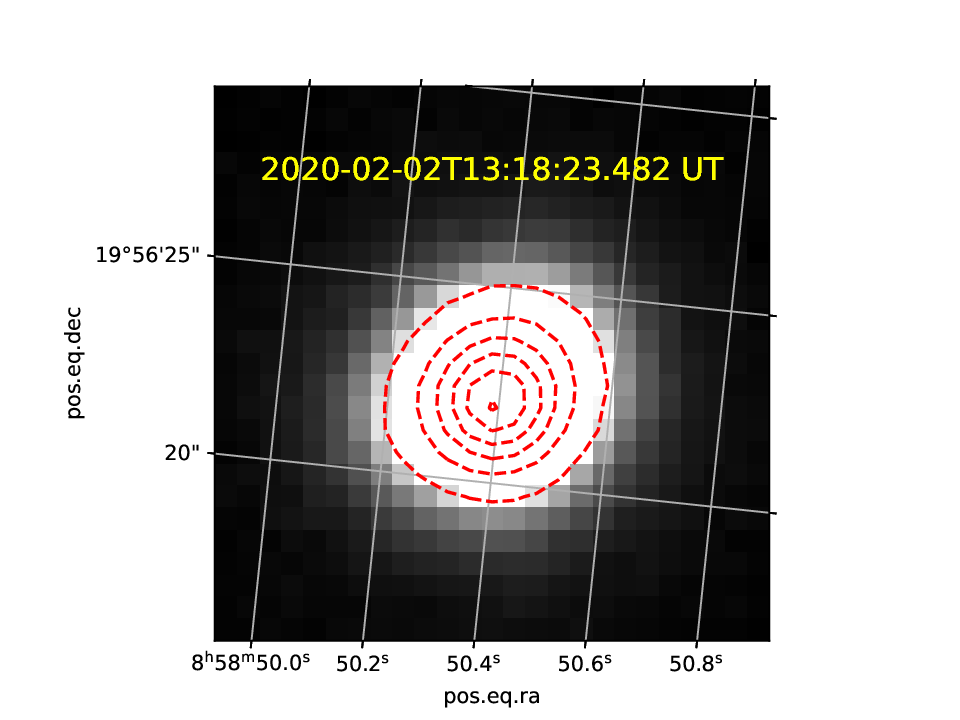} 
   \includegraphics[width=0.4\textwidth]{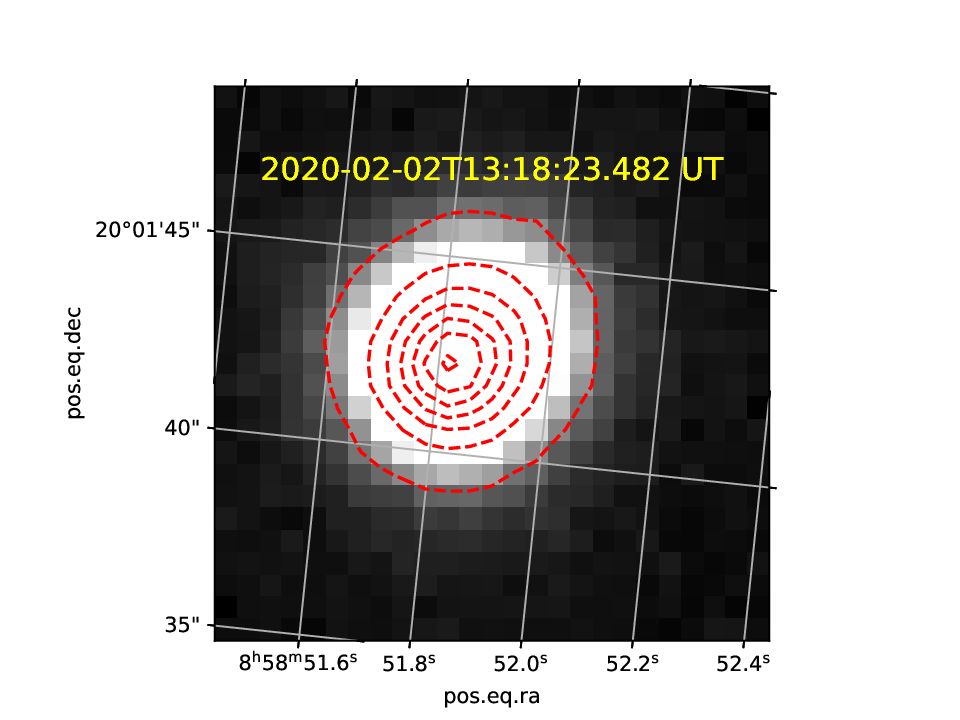}
       \includegraphics[width=0.4\textwidth]{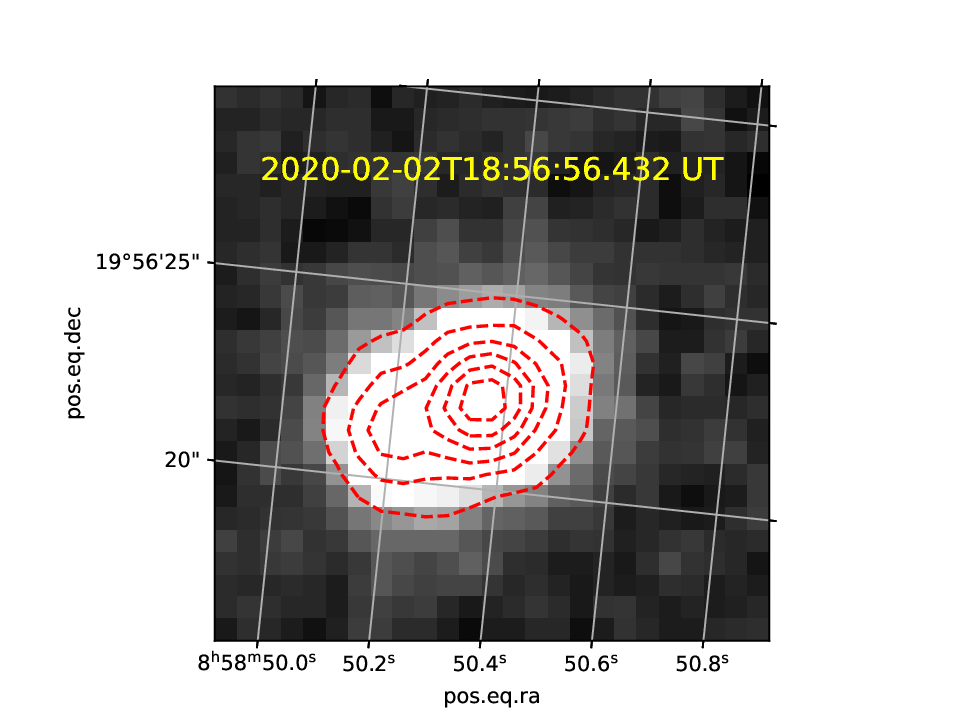} 
   \includegraphics[width=0.4\textwidth]{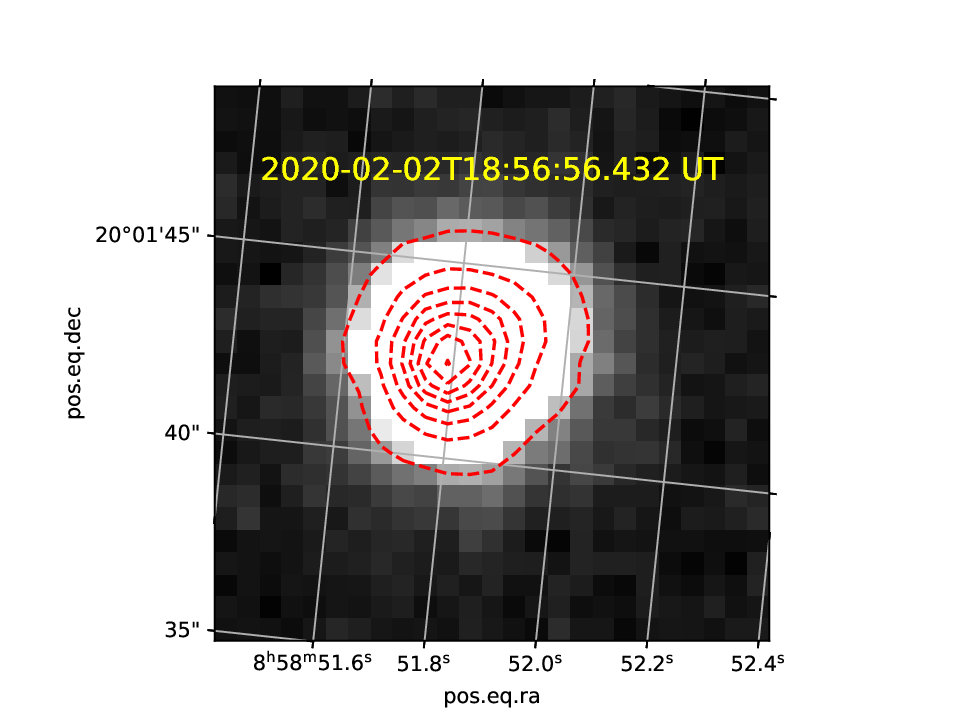}
   \caption{The 25 $\times$ 25 pixels $R-$band images of the field of EI~Cnc 
   obtained by the F60A telescope in 2020-02-02 (the original size of the image is 19.5 $\times$ 19.5 arcmin$^{2}$). The left and right columns correspond to the object and the comparison star, respectively.
   The upper row shows the first image obtained at 2020-02-02UT13:18:23.482,
   while the bottom rows the last image at UT18:57:20.759, about 5.5 hours later than the first one.}
   \label{fig:F60Aobs}
   \end{figure*}

\subsection{Data process}

The images taken by both GWAC FFoV and F60A telescope were processed by the standard procedure, including bias, dark, and flat-field corrections, using the IRAF\footnote{IRAF is distributed by the National Optical Astronomical Observatories, which are operated by the Association of Universities for Research in Astronomy, Inc., under cooperative agreement with the National Science Foundation.} package.
In the subsequent aperture photometry,
the aperture was adopted to be 3.0 and 10.0 pixels for the images taken by the FFoV and F60A, respectively. 
In order to combine the light curves resulted from FFoV and F60A,
we used a relatively large aperture for the F60A images to enclose both components 
of the object, both because the two components are marginally
resolved in the F60A images even at the quiescent state and because of
the low spatial resolution of the FFoV images. 
Absolute photometric calibration was performed using the
USNO B1.0 catalogure \citep{Monet2003}.
The comparison stars used in our differential 
photometry are tabulated in Table \ref{Tab2}.
 
\begin{table}[ht]
   \caption{The comparison stars used to calibrate the light curve of EI~Cnc. 
   The $R2$ and $B2$ magnitudes are derived from the USNO B1.0 catalog \citep{Monet2003}. }
   \centering
   \begin{tabular}{lllll}
   \hline
  GWAC camera    & R.A.   & Decl.    &  $R2$  & $B2$ \\
or telescope &(J2000) & (J2000) & mag & mag \\
\hline
    GWAC/FFoV   & 134.715109  & 19.800973 & 9.19 & 9.89 \\
    F60A  & 134.564123  & 19.851556 & 11.62 & 11.09 \\
   \hline
 
    \end{tabular}
    \label{Tab2}
   \end{table}

\subsection{Determine the source of the superflare}

Figure.\ref{fig:F60Aobs} shows the brightness contour of the $R-$band images obtained 
by F60A at two epochs. According to our astrometry, the contour was centered at EI~CncA near the peak of the flare. While, the contour shape was lengthened by the contribution of EI~CncB 
as the flare faded out, although the brightness was still dominated by EI~CncA.
This analysis 
enables us to claim that the superflare occurring in 2020-02-02 is produced by the brighter component EI~CncA, thanks for the rapid follow-ups by F60A with relatively higher spatial resolution of 0.52 arcseconds per pixel.

\subsection{Properties of the light curve}

\begin{figure*}[htbp]
    \centering
   \includegraphics[width=0.8\textwidth]{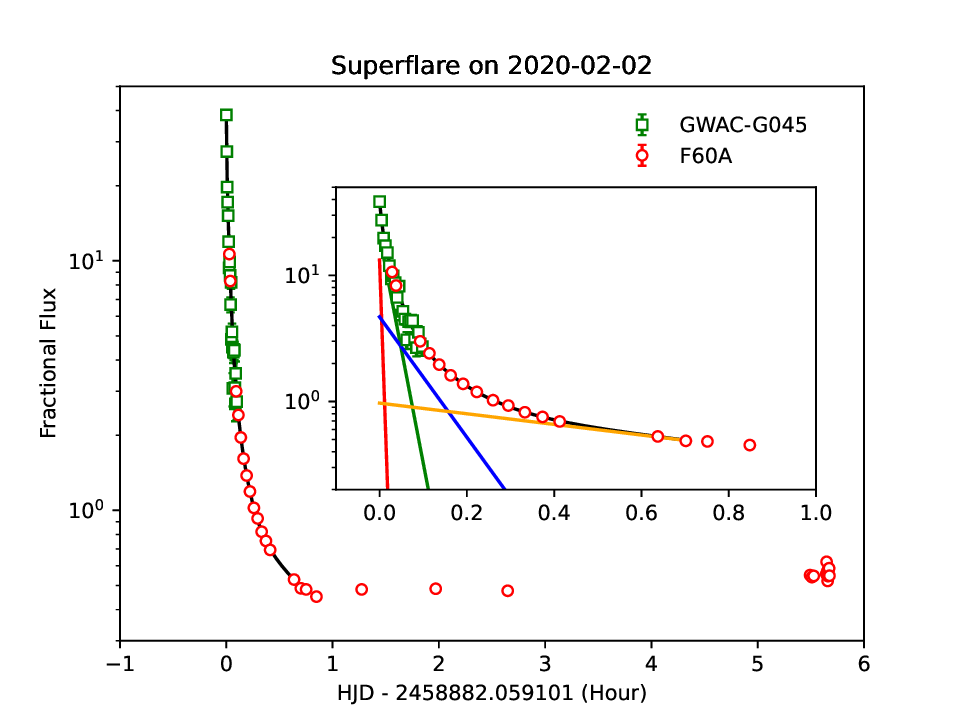} 
   \caption{Combined $R$-band light curve of the superflare tiggered by the GWAC/FFoV in 2020-02-02. The green squares refers to the data obtained by the FFoV, and the red circles the data by F60A. The black curve shows the best fit to the light curve. The four components 
   used to reproduce the light curve are shown separately by different colors in the insert sub-panel.  }
   \label{fig:superflare}
   \end{figure*}

Figure.\ref{fig:superflare} shows the temporal variation of the calculated fractional flux in the $R-$band of the superflare. 
The fractional flux is calculated by adopting a quiescent specific flux of 
$F_{R,q}$ = 1.24 $\times$ $10^{-14}$  erg cm$^{-2}$ s$^{-1}$ \AA$^{-1}$
that is transformed from the quiescent brightness of $R=12.87$ mag extracted from the USNO B1.0 catalogue.

Although the sky field where the EI~Cnc locates in has been monitored by FFoV long enough 
before the 
flare occurred, the rising phase of the superflare was not recorded, which suggests a quite fast rising with a rate of $>0.16\ \mathrm{mag\ s^{-1}}$, 
if the corresponding limit magnitude is used a reference. By adopting 
quiescent brightness is adopted as a reference, 
the rising rate is expected to be less than $0.24\ \mathrm{mag\ s^{-1}}$.

With the high cadence of 15 seconds, we are able to study the
decay behavior of the superflare in more details by modeling the 
decay phase with the following formula: 

\begin{equation} 
\frac{F_{\text {decay}}}{F_{\text{amp}}} = \sum_{i=1}^{n} a_i e^{-k_i \times t/t_{1/2} }   
\label{func1}       
\end{equation}
where $F_{\mathrm{amp}}$ is the peak flux of the light curve, $a$ the amplitude of the component, $k$ the decaying index for each component, and $t$
the time after the trigger time. $t_{1/2}$ is the decay duration from the 
peak to the half of the peak. The Bayesian information criterion (BIC) is adopted in
the modeling to test whether the fitting is overfitted or not. 
The BIC value is 137.33 for the model with 3 components, 
while 75.46 for the model with 4 components.
The result indicates that the 4-component model is the best one 
with a reduced $\chi^2/\mathrm{d.o.f.}=13.10$ and a degree of freedom of 47.
The best fit model is shown in Fig.\ref{fig:superflare} by different colors. The values of the corresponding 
parameters are tabulated in the Table \ref{Tab3}.

During the modeling of the superflare, only the peak time to 2500 seconds of the light curve was analysed.
The brightness after 2500 seconds was not integrated since the later light curve had become too shallow which was likely not part of the flare, but might be caused by the variation of the object itself.

Given the best fit model, the equivalent duration time ($ED$), the time needed
to emit the flare energy at the quiescent level, is estimated to be $5293$ seconds or $\sim$1.5 hours.
With the distance of 5.13 pc and the determined $ED$, the lower limit of total $R$-band flare energy $E_{R}$ is inferred to be $E_{R} = (3.3\pm0.2) \times 10^{32} $  ergs.

\begin{table}[ht]
   \caption{The parameters of the four-components model resulted from our light curve fitting.
   $k_i$ and $\alpha_i$ are the decay slope and 
   normalization of each component, respectively.  }
   \centering
   \begin{tabular}{cc}
   \hline\hline
   Parameters & Value  \\
   \hline 
   $a_{1}$  &  0.508869  $\pm$ 0.007355\\
   $k_{1}$  &  0.011504  $\pm$ 0.000241\\
   $a_{2}$  &  0.310620  $\pm$ 0.022111 \\
   $k_{2}$  &  0.062431  $\pm$ 0.008285\\ 
   $a_{3}$  &  0.116739  $\pm$ 0.003715  \\
   $k_{3}$  &  0.003050  $\pm$ 0.000073 \\ 
   $a_{4}$  &  0.024246  $\pm$ 0.000554  \\
   $k_{4}$  &  0.000270  $\pm$ 0.000010 \\ 
    \hline
    \end{tabular}
   \label{Tab3}
   \end{table}

\section{Off-line Search for Flares from GWAC Archive images}

\begin{figure}[htbp]
    \centering
   \includegraphics[width=0.4\textwidth]{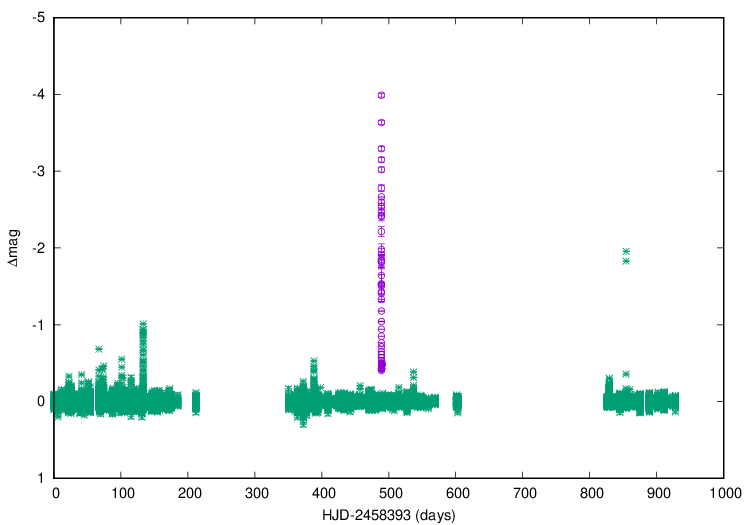} 
   \caption{The long-term light curve of EI~Cnc obtained by GWAC JFoV after subtracting the quiescent flux level. The super flare on 2020-02-02 is marked by the magenta circles.}
   \label{fig:alllc}
\end{figure}

The on-line pipeline we have developed is intend to detect large-amplitude superflares in real-time \citep{Xin2021b}. 
In order to study the flare activities of EI~Cnc in a long term,
we performed an off-line search for flares of EI~Cnc, which 
might be missed by our on-line pipeline,
in the JFoV archive images obtained from 2018 October 1 to 2021 April 24. 
By using the data reduction methods described in Section 3.2,
the long-term light curve obtained by the JFoV is shown in Figure \ref{fig:alllc}.
In total, the object has been recorded in 78422 images over 198 nights, spanning $\sim$290 hours.
Considering the fact that the flare duration is 
typical of tens of minutes to a few hours, our off-line search is designed as follows.

\begin{enumerate}
    \item The light curve in each night is produced by a differential photometry with 
    the same companion star for all the images. 
    \item Each generated light curve is sliced into several boxes, each with a time window
    of 30 minutes.
    \item The medium and standard derivation ($\sigma$) are calculated in each box.
    \item Any measurement with a deviation greater than $+3\sigma$ is marked. 
    \item A flare event candidate is flagged if there are more than three consecutively marked measurements.
    \item All the candidates are examined by human eyes one by one. 
\end{enumerate}

As a result, in addition the superflare described in Section 3, a total of 27 flares were identified within the 2018-2021 observing seasons. The log of all the flares is shown in 
Table \ref{Tab4}, in which the peak time 
and flare amplitude are listed in Columns (2) and (3), respectively.
Figure.\ref{fig:flares_GWAC} and Figure.\ref{fig:flares_GWAC_2}
display the zoom-in light curves of the 27 flares.

One can see from the figures 
that five (20$\%$) out of the 27 flares 
show complex eruptions with multi peaks.
This fraction is higher than $12\%$ found in the TRAPPIST- sample \citep{Vida2017}, simply because of the GWAC's high cadence of 
15 seconds. Based on the $K2$ measurements, the cadence is $\sim$59s for the TRAPPIST-1 sample.

The energies released in these flares are estimated by the same method
previously described in Section 3.3. 
Lower limits of the released energies are calculated for 4 out of the 27 flares 
without complete light curves. 
The measured flare amplitude and estimated $E_R$ are tabulated in Columns (3)  and (5) in Table \ref{Tab4}, respectively.
The flare amplitude ranges from $\Delta R\sim0.1$ mag to
$\sim$3 mag, which corresponds to a
flare energy range between $E_R=4.0 \times 10^{29}$ erg and $3.3 \times 10^{32}$ erg.

\begin{figure*}[htbp]
   \centering
      \includegraphics[width=0.9\textwidth]{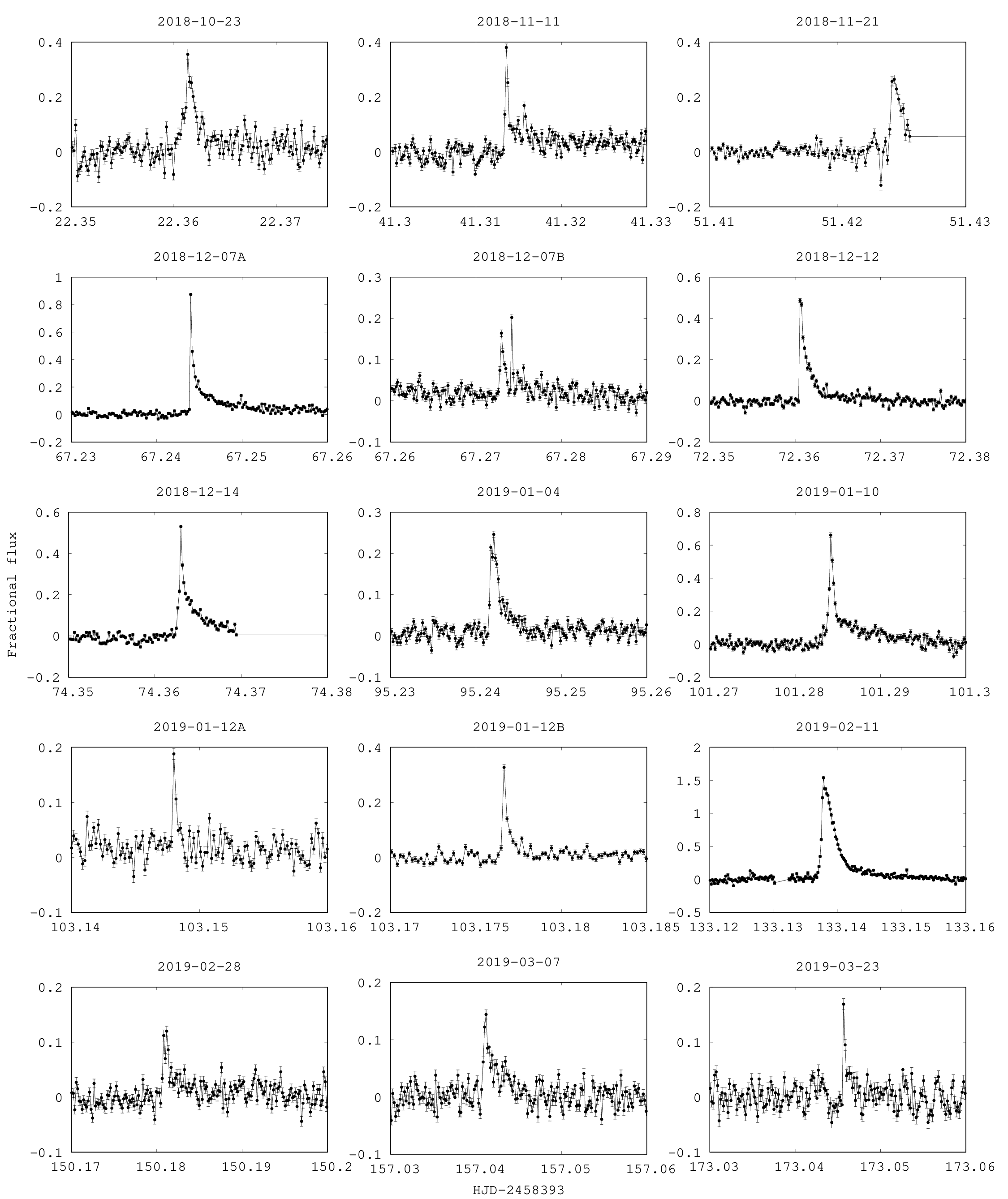} 
         \caption{The  $R-$band light curves of the 27 flares 
         found by our off-line search in the GWAC JFoV
         achieved data obtained from 2018 October 1 to 2021 April 24. 
         In order to display the main flare phase better, the tails of some of the  
         flares are not shown due to a cutoff on the abscissa axis.}

   \label{fig:flares_GWAC}
   \end{figure*}

\begin{figure*}[htbp]
   \centering
      \includegraphics[width=0.9\textwidth]{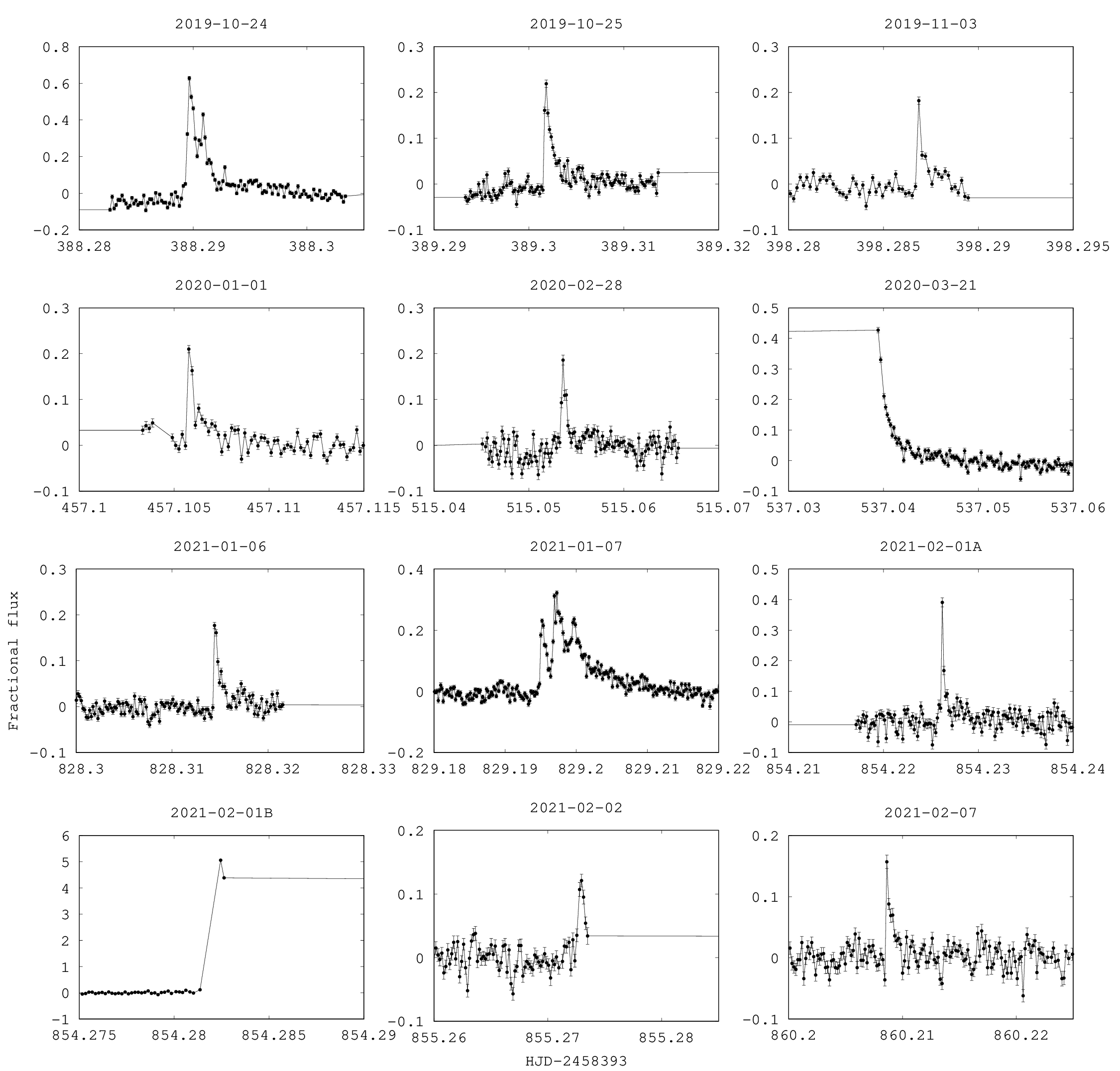}
   \caption{Continued for the Figure.\ref{fig:flares_GWAC}}
   \label{fig:flares_GWAC_2}
   \end{figure*}

\begin{table*}[htbp]
\caption{Properties of all the 28 flares of EI~Cnc detected in this study. 
}
\centering
\setlength{\tabcolsep}{1mm}{
\begin{tabular}{cccccc}
\hline\hline
No. &  Peak time  & $\Delta R$  &    $\tau$  & $E_R$        &   $B$\\
   &      UT       &  mag        &     second    &  erg     &  Gauss\\
(1) &  (2)  & (3) & (4) & (5) & (6)\\ 
\hline
1  &  2018-10-23T20:41:53 & $0.34\pm0.02$  & 76  & $8.6\times10^{30}$  &  161 \\
2  &  2018-11-11T19:30:20 & $0.36\pm0.01$  & 30 & $1.0\times10^{31}$   &  174 \\
3  &  2018-11-21T22:08:32 & $0.25\pm0.02$  & 82 & $>4.6\times10^{30}$  &  $>$118 \\
4  &  2018-12-07T17:46:40 & $0.68\pm0.01$  & 33 & $2.2\times10^{31}$   &  258 \\
5  &  2018-12-07T18:30:10 & $0.20\pm0.01$  & 43 & $4.1\times10^{30}$   &  111 \\
6  &  2018-12-12T20:33:58 & $0.44\pm0.01$  & 81 & $9.7\times10^{30}$   &  171 \\
7  &  2018-12-14T20:37:16 & $0.48\pm0.01$  & 59 & $1.4\times10^{31}$   &  206 \\
8  &  2019-01-04T17:41:17 & $0.24\pm0.01$  & 77 & $6.0\times10^{30}$   &  135\\
9  &  2019-01-10T18:41:31 & $0.55\pm0.01$  & 50 & $1.2\times10^{31}$   &  191\\
10 &  2019-01-12T15:25:20 & $0.15\pm0.01$  & 17 & $4.0\times10^{29}$   &  35\\
11 &  2019-01-12T16:06:35 & $0.27\pm0.01$  & 17 & $6.6\times10^{29}$   &  45\\
12 &  2019-02-11T15:10:14 & $1.03\pm0.01$  & 187 & $6.8\times10^{31}$  &  454\\
13 &  2019-02-28T16:13:32 & $0.12\pm0.01$  & 50 & $1.2\times10^{30}$   &  60\\
14 &  2019-03-07T12:52:28 & $0.15\pm0.01$  & 73 & $2.9\times10^{30}$   &  94\\
15 &  2019-03-23T13:00:28 & $0.17\pm0.01$  & 21 & $1.8\times10^{30}$   &  74\\ 
16 &  2019-10-24T18:58:32 & $0.58\pm0.01$  & 118 & $3.2\times10^{31}$  &  311\\
17 &  2019-10-25T19:15:55 & $0.22\pm0.01$  & 61 & $5.0\times10^{30}$   &  123\\
18 &  2019-11-03T18:53:04 & $0.18\pm0.01$  & 15 & $6.9\times10^{29}$   &  46\\
19 &  2020-01-01T14:25:13  & $0.20\pm0.01$ & 24 & $1.6\times10^{30}$   &  70\\
20 &  2020-02-02T13:16:50 & $3.70\pm0.01$  & 73 & $3.3\times10^{32}$   &  1000\\
21 &  2020-02-28T13:09:49 & $0.20\pm0.01$  & 38 & $3.3\times10^{30}$   &  100\\
22 &  2020-03-21T12:51:21 & $>0.40\pm0.01$ & 86 & $>7.4\times10^{30}$  &  $>$150\\
23 &  2021-01-06T19:25:31 & $0.18\pm0.01$  & 55 & $1.8\times10^{30}$   &  74\\
24 &  2021-01-07T16:36:01 & $0.32\pm0.01$  & 334 & $1.9\times10^{31}$  &  240 \\
25 &  2021-02-01T17:17:27 & $0.36\pm0.01$  & 19 & $2.4\times10^{30}$   &  85\\
26 &  2021-02-01T18:37:23 & $>2.29\pm0.01$ & 182 & $>2.1\times10^{32}$ &  $>$798\\
27 &  2021-02-02T18:24:48 & $0.12\pm0.01$  & 38 & $>1.2\times10^{30}$  &  $>$60\\
28 &  2021-02-07T16:52:13  & $0.15\pm0.01$ & 52 & $9.8\times10^{30}$   &  172 \\
 \hline
 \end{tabular}}
\label{Tab4}
\end{table*}

\section{Cumulative Flare Energy Distribution}

\begin{figure}[htbp]
   \centering
   \includegraphics[width=0.45\textwidth]{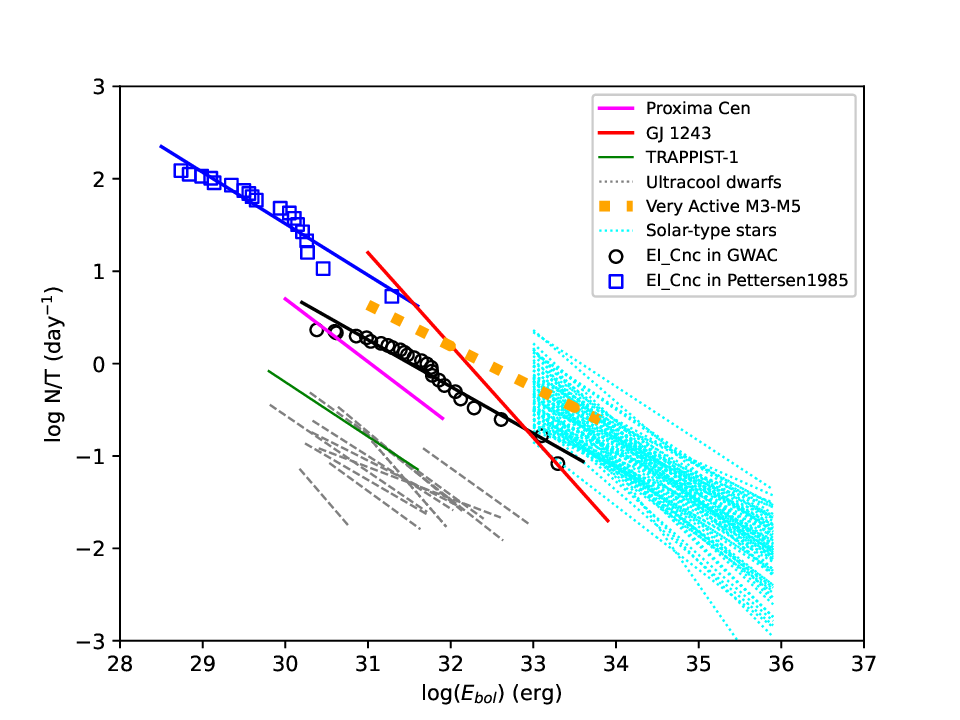} 
   \caption{A comparison of the cumulative FFD.
   The FFDs inferred from this study and reported in \citet{Pettersen1985b} are 
   presented by the black-open circles and by the blue-open squares, respectively.
   The black and blue solid lines denote the best fit linear relationships  
   according to Eq. (2). 
   The magenta, red and green solid lines represent of the aproximate positions of FFDs of Proxima Cen, GJ 1243 and TRAPPIST-1 \citep{Vida2017} respectively.  
   The grey dashed lines and orange dotted line represent FFDs of ultracool dwarfs from \citet{Paudel2018} and very active M3-M5 dwarfs \citep{Hawley2014} respectively.
   The cyan dotted lines refer to FFDs of solar-type stars \citep{Gao2022}.}
   \label{fig:ffd4eicnc}
   \end{figure}

Based on the flares reported above, we investigate the cumulative
flare frequency distribution (FFD) of EI~Cnc in this section. 
After excluding the flares with only a lower limit of $E_R$,
the calculated FFD is plotted as a function of bolometric flaring energy 
$E_{\mathrm{bol}}$ in Figure \ref{fig:ffd4eicnc}.  
$E_{\mathrm{bol}}$ is converted from $E_R$ 
by a bolometric correction of $E_{\mathrm{bol}}=E_R\times6.0$ assuming 10,000K blackbody radiation
\citep[e.g.,][]{Gizis2013, Kowalski2013,Paudel2019,Murray2022,Fleming2022}.
The resulted FFD shows a clear deviation from the linear fit both at low and high energies. 
Similar feature could also be seen in other active stars, such as GJ\,1245AB \citep{Lurie2015} and GJ\,1243 \citep{Silverberg2016}.
One explanation for this behavior is the incomplete detection of low and high energy flares, but the possibility of a real deviation cannot be excluded \citep{Hawley2014,Lurie2015}. More intensive observations are needed to confirm this.
Following the commonly used method, we fit the FFD by a linear function of 

\begin{equation}
\log \widetilde{\nu} =\alpha + \beta\log {E_{\mathrm{bol}}} 
\end{equation}

where $\widetilde{\nu}$ and $E_{\mathrm{bol}}$ are the cumulative flare frequency per day and 
the bolometric flare energy in unit of ergs, respectively. $\alpha$ and $\beta$ are the normalization and distribution index, respectively.
By assuming that the uncertainty of FFD could be described by a Poisson sampling  following \citet{Lurie2015},
a least-square fit returns $\alpha=15.26\pm 0.96$ and $\beta=-0.50\pm0.03$, along with 
a reduced $\chi^2$ of 0.14 at a degree of freedom of 20.
The best-fit relationship is overplotted in Figure.\ref{fig:ffd4eicnc} by 
a black line. 

The distribution index $\beta$ is traditionally used to characterise the flare energy distributions. For stars with $\beta>-1$, high-energy flares dominate the total energy output released in the flaring process \citep{Paudel2018, Jackman2021}.
However, the total energy output is dominated by low-energy flares for the 
stars with $\beta<-1$, which could be understood by the heating of the quiescent corona \citep{Parker1988, Hudson1991, Schrijver2012}.   
Based on the GWAC observations, the modeled $\beta\sim-0.5$
suggests a quite shallow FFD for EI~Cnc, a M7 type dwarf. 
This shallow FFD reinforces the trend reported recently in 
a comprehensive study on the Kepler and TESS photometric data by
\citet{Gao2022}, in which, compared to early M-dwarfs, 
stars cooler than M4 show marginally enhanced superflare activity.

Similar to those found in a lot of previous studies \citep[e.g.][]{Lurie2015,Silverberg2016,Vida2017, Paudel2018,Lin2022},
a close inspection of the resulted FFD shows an obvious deviation from a 
single powerlaw by a ''smooth'' break at 
$E_{\mathrm{bol}}\sim10^{32}\ \mathrm{erg}$.
Although an incompleteness of flares can not be entirely excluded,
further study is needed to confirm the phenomenon and to understand the 
potential underlying physics.  

\section{Discussion}

\subsection{A comparison study on FFD}

The blue-open squares, along with the blue solid line, 
show the FFD of EI~Cnc generated from \citet{Pettersen1985b} who
reported 13 flares during 4.5 hours in the $U$-band. 
In order to compare the FFDs in different epochs, 
the $U$-band flaring energy $E_U$ is in advance converted to 
$E_{\mathrm{bol}}$ by adopting a bolometric correction of $E_{\mathrm{bol}}=E_U\times6.0 $, after 
assuming a blackbody with a temperature of 10,000K at the peak time
\citep{Kowalski2013}. 

Comparing this FFD measured about four decades ago with the one given in this study enables 
us to reveal a long term variation of the flare rate and magnetic activity of EI~Cnc. 
The extrapolation is reasonable after taking into account of the studies of the 
flare activity in a number of the solar-type and cooler active stars (e.g., \citet{Maehara2012,Paudel2018,Gao2022}), 
in which a global FFD slope is available for most of the 
stars in a wide flare energy ranging from
the nano-flares ($\sim10^{24}$ erg) to the superflares ($\sim10^{36}$ erg).

On the other hand, at energy of $\sim10^{31}\ \mathrm{erg}$, the FFD from 
\citet{Pettersen1985b} yields a flare rate of about ten times per day, which is  
quite close to those of active M6 dwarf Wolf~359 \citep{Lin2022} and 
very active M3-M5 dwarfs (the orange dots in Figure \ref{fig:ffd4eicnc}) as reported by \citet{Hawley2014}. 

Figure \ref{fig:ffd4eicnc} additionally compares the FFD of EI~Cnc measured in the 
current study with those of other M dwarfs after adopting a relationship of $E_{\mathrm{bol}}=3.1E_{\mathrm{Kp}}$ \citep{Paudel2018}, where $E_{\mathrm{Kp}}$ is the 
\it Kepler \rm flare energy. 
The comparison shows that EI~Cnc is in fact 
more active than not only 
the M5.5 dwarf Proxima Cen (\cite{Davenport2016}, the magenta line), but also 
the ultracool dwarfs (the grey dashed lines) studied in \citet{Paudel2018}. 

Especially, the flare rate of EI~Cnc is found to be at least ten times higher 
than that of the M8 dwarf TRAPPIST-1 (the green line, \citet{Vida2017}).
Further more, compared to EI~Cnc, a much steeper FFD is found for the active 
M4 dwarf GJ~1243.
If the flaring rate of EI~Cnc is extrapolated to the energy larger than $10^{34}$ erg from the fit of FFD,  then its superflare rate with energies E$>10^{34}$ erg would be at least one order of magnitude higher than that of GJ~1243.
The reason for high superflare rate is still under debate at the current stage.
There are some evidences supporting that superflares are more likely to occur 
on young stars with fast rotation \citep{Howard2020}. However, 
other studies found an enhanced superflare rate in stars  
with intermediate rotation periods (10-70d)\citep{Mondrik2019}.

The FFDs of solar-types stars \citep{Gao2022} are  collected and displayed in the Figure \ref{fig:ffd4eicnc} as cyan dotted lines.  The data from solar-type stars have a wide distribution indicating a different level of activity. The slopes of these FFDs are similar with each other, and also comparable to the slopes of EI~Cnc we derived, suggesting similar flare generation mechanism for stars from M-types to solar types.

There has been a few attempts made to explore long-term variations of cool dwarfs (eg., \citet{Buccino2011,Silva2012,Bustos2020,Mignon2023}).
In the aspect of flaring rate, \citet{Davenport2020} reported no sign of solar-like activity cycles 
for the active flaring M4 dwarf GJ~1243 over 10 years.
Lately, \citet{Crowley2022} found a low number of stars with detectable rate variation 
by analysing TESS data of 274 G type stars.
For the case of  EI~Cnc, although the two FFDs have small common energy range, the current FFD is 5$\sim$6 times 
lower than the extrapolation of the FFD generated from \citet{Pettersen1985b},
which may be related to the active cycles in ultra cool dwarfs, 
similar to those occurred on the Sun \citep{Usoskin2023} and solar type stars \citep{Baliunas1995,Berdyugina2005,Olspert2018,Saikia2018,Baum2022}. 

\subsection{The flare duration verses energy}

\citet{Maehara2015} reports a correlation between white light (WL) flare energy and duration 
$\tau$ for the superflares of solar-like stars: $\tau\propto E^{0.39\pm0.03}$, where $\tau$ 
is defined as the $e-$folding decay time of flare intensity after its peak and E is defined as the bolometric energy of flares. 

On the observational ground, this correlation is quite similar with not only the
relationship of $\tau\propto E^{0.2-0.3}$ in the solar hard/soft X-ray flares (e.g., 
\citet{Veronig2002}, \citet{Christe2008}), but also the relationship of $\tau\propto E^{0.39}$
revealed in the solar WL flares \citep{Namekata2017}. In addition, the validation of the 
$\tau-E$ relationship has been extended to mid-M dwarfs by a comprehensive study carried 
out by Chang et al. (2015). Such similarity therefore reinforces the ideal that
solar and stellar flares share the same mechanism: 
the magnetic reconnection, although there is some 
difference in the relationship between the field mid-M dwarfs and the mid-M stars in open 
cluster \citep{Chang2015}. 
On the theoretical ground, a $\tau\propto E^{1/3}$ relationship could be obtained in
the magnetic reconnection scenario, if both magnetic field strength $B$ and Alfven velocity
$\upsilon_{\rm A}=B/\sqrt{4\pi\rho}$ are taken to be constants \citep{Maehara2015}.

\begin{figure}[htbp]
   \centering
   \includegraphics[width=0.5\textwidth]{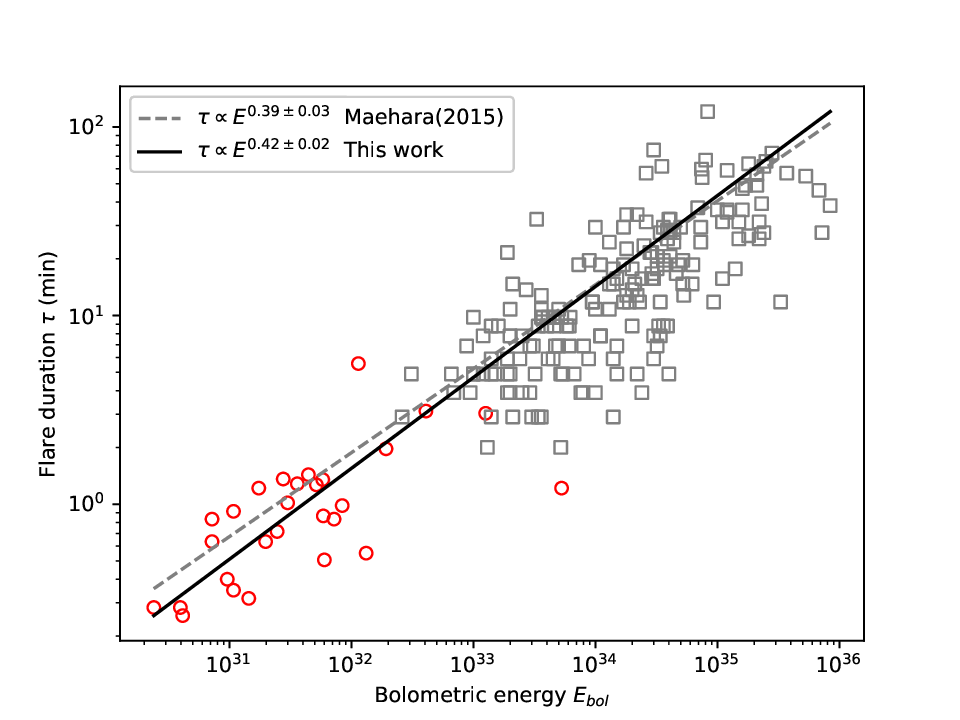} 
   \caption{The flare duration $e$-folding time $\tau$ verses the bolometric flare energy. 
            The red circles and black squares denote the flares detected in M7 dwarf EI~Cnc 
            in this work and the superflares on G-type stars detected from the Kepler 
            short cadence data by \citet{Maehara2015}, respectively. 
            The dashed line shows the best fit
            $\tau-E$ relationship given in \citet{Maehara2015}, when the superflares are considered. 
            After including the flares in EI~Cn, an updated $\tau-E$ relationship
            is overplotted by the solid line.}
   \label{fig:e_tao}
   \end{figure}

Following \citet{Maehara2015}, we show the $\tau-E$ relationship for the
WL flares detected in EI~Cnc by the red circles in Figure.\ref{fig:e_tao}. 
The measured $\tau$ of each of the flares are listed in Column (4) in Table \ref{Tab4}. 
$E$ is obtained from the measured $E_R$ by a transform used by \citet{Maehara2015}.  
In contrast to the results of the Sun given in \citet{Namekata2017}, 
the figure shows that the WL flares 
detected in EI~Cnc closely follow the $\tau-E$ relationship established in the WL 
superflares of the solar-type stars by extending the flare energy down to 
$\sim10^{30}$erg, which 
suggests that the late-M dwarfs and the partially convective solar-type stars 
share a common flare mechanism. Combining the two data sets shown in the figure returns a best-fit $\tau-E$ relationship 

\begin{equation}
    \tau \propto E^{0.42 \pm 0.02}
\label{eq:e_tao}
\end{equation}
which is valid for the flares of both late-M dwarfs and solar-type stars.

Assuming $\tau$ is comparable to the reconnection timescale, 
the complete scaling law can be expressed as 
$\tau\propto E^{1/3}B^{-2/3}\upsilon_{\rm A}^{-1}M_{\rm A}^{-1}$ \citep{Namekata2017}, 
where $\tau$ and E refer to the $e-$folding decay time and the bolometric energy of flares respectively, $\upsilon_{\rm A}$ is the Alfv\'en velocity, and $M_{\rm A}$ is the dimensionless reconnection rate ranging from 0.1 to 0.01 in the Petschek-type fast reconnection 
\citep[e.g.,][]{Shibata2011}. Taking into account of the considerable difference in
$B$ and $\rho$ between late-M dwarfs and solar-type stars, the universal $\tau-E$ relationship 
shown in the figure implies $B^{2/3}\upsilon_{\rm A}$ is an invariant in the (super)WL flares 
on stars with different types.

\subsection{Estimation of magnetic field}

Based on the solar flare model, the magnetic field strength $B$ of EI~Cnc can be 
estimated from flare energy $E_{\mathrm{flare}}$ according to \citep{Shibata2013}:

\begin{equation} 
   \begin{aligned}
    E_{\mathrm{flare}} \approx 7\times10^{32}(\textrm{erg})\bigg(\frac{f}{0.1}\bigg)\bigg(\frac{B}{10^{3}\textrm{G}}\bigg)^{2}\bigg(\frac{A_{\textrm{spot}}}{3\times10^{19}\textrm{cm}^{2}}\bigg)^{3/2}\ \mathrm{erg}
\end{aligned}
\label{eq:shibata2013rev}
\end{equation}
where $f$ is the fraction of magnetic energy that can be released as flare energy.
$A_{\rm spot}=4\pi R_*^2f_{\rm spot}$ is the area of spot on a star, 
where $R_*$ is the stellar radius and $f_{\rm spot}<1$ the fraction of stellar area.  

The value of $B$ of each of the flares identified in this study
is estimated by adopting a typical value of $f=0.1$ \citep{Aschwanden2014,Jackman2018} 
and the estimated stellar radius of $R_* \sim 0.2R\odot$ for EI~Cnc \citep{Gershberg1999},
and listed in column (6) in Table \ref{Tab4}.
A fiducial value of $f_{\rm spot}=0.1$ is used in the estimations. 
The estimated $B$ has a value of $\sim10^3$Gauss for the energetic 
flares with $E\sim10^{33}$erg, which is comparable to
those estimated for other superflares of later M dwarfs 
\citep[e.g.,][]{Paudel2018,Xin2021b}.

\section{Summary}

An optical monitoring campaign on the nearby flaring system EI Cnc was 
carried out by the GWAC and its dedicated follow-up telescope, which allows us to 
arrived at following results.
\begin{enumerate}
    \item A superflare with a lower limit of R band flare energy $\sim 3.3\times10^{32}$erg was 
    triggered and observed, in which four components are required to properly model the 
    complex decay light curve. Rapid follow-ups revealed that the flare was generated by the brighter component of the binary EI~CncA.
    \item Combining the 27 flares additionally detected from the GWAC achieve data leads to 
    a quite shallow 
    FFD with $\beta=-0.50\pm0.03$ over an energy range between $10^{30}$erg 
    and $10^{33}$erg, which supports an dominance of high-energy flares in the  
    the flare activity in EI Cnc.
    \item The significant decreasing flare rate is obtained compared with the observations about four decades ago, which is probably related to the active cycles in ultra cool dwarfs.
    \item The white-light (super)flares from late M dwarf and solar-type stars are found to follow an updated $\tau-E$ relationship $\tau\propto E^{0.42\pm0.02}$, which implies a universal mechanism in the stellar flare activity. 
    
\end{enumerate}

\section{Acknowledgement}

We thank the anonymous referee for helpful comments that allow us to improve the paper significantly. 
This study is supported from the National Natural Science Foundation of China (Grant No. 11973055, U1938201, U1831207, U1931133,12133003) and partially supported by the Strategic Pioneer
Program on Space Science, Chinese Academy of Sciences, grant
Nos. XDA15052600 and XDA15016500. 
JW is supported by the National Natural Science Foundation of China under grants 12173009 and by the Natural Science. 
Foundation of Guangxi (2020GXNSFDA238018).
DWX is supported by the National Natural Science Foundation of 
China under grant 12273054.
YGY is supported by the National Natural Science Foundation of China under grant 11873003. 
HBC is supported by the National Natural Science Foundation of China under grant 11973063.
This research has made use of the VizieR catalogue access tool, CDS, Strasbourg, France (DOI: 10.26093/cds/vizier). 
The original description of the VizieR service was published in A\&AS 143, 23. This work has made use of data from the European Space Agency (ESA) mission
{\it Gaia} (\url{https://www.cosmos.esa.int/gaia}), processed by the {\it Gaia} Data Processing and Analysis Consortium (DPAC,
\url{https://www.cosmos.esa.int/web/gaia/dpac/consortium}). Funding for the DPAC
has been provided by national institutions, in particular the institutions
participating in the {\it Gaia} Multilateral Agreement.

\vspace{5mm}
\facilities{GWAC, GWAC-F60 telescope}

\software{IRAF (Tody 1986, 1992)
          }

\end{document}